\documentclass[
letter,
reprint,
amsmath,amssymb,aps
]{revtex4-1}

\usepackage[utf8]{inputenc}
\usepackage{amsmath,amsfonts,amstext,amssymb,mathrsfs,amsthm}
\usepackage{graphicx}
\usepackage{hyperref}
\usepackage{xcolor}
\usepackage{tikz}

\usetikzlibrary{decorations.pathmorphing}
\tikzset{snake it/.style={decorate, decoration=snake}}

\hypersetup{
    colorlinks,
    linkcolor={blue!80!black},
    citecolor={blue!80!black},
    urlcolor={blue!40!black}
}

\theoremstyle{theorem}
\newtheorem{claim}{Claim}

\newcommand{\D}{\mathrm{d}}
\renewcommand{\O}{\mathcal{O}}
\newcommand{\<}{\langle}
\renewcommand{\>}{\rangle}

\begin{document}

\title{Breaking the spell of the tensor product}
\author{Adam Bzowski}
\affiliation{Crete Center for Theoretical Physics, 
	Department of Physics, 
	University of Crete, 
	70013 Heraklion, 
	Greece\\
email: a.bzowski@physics.uoc.gr}
\date{\today}

\begin{abstract}

\noindent In this essay I argue that the Hilbert space of states dual to a traversable wormhole is smaller than the tensor product of the independent Hilbert spaces of the boundary field theories. From the point of view of semiclassical physics, the decrease in the number of states is perceived as an emergent, non-local interaction stabilizing the wormhole. This presents new possibilities for models of radiating black holes and raises questions about  results established under the spell of the tensor product.

\bigskip

\noindent \textit{Essay written for the Gravity Research Foundation 2021 Awards for Essays on Gravitation}

\end{abstract}

\maketitle

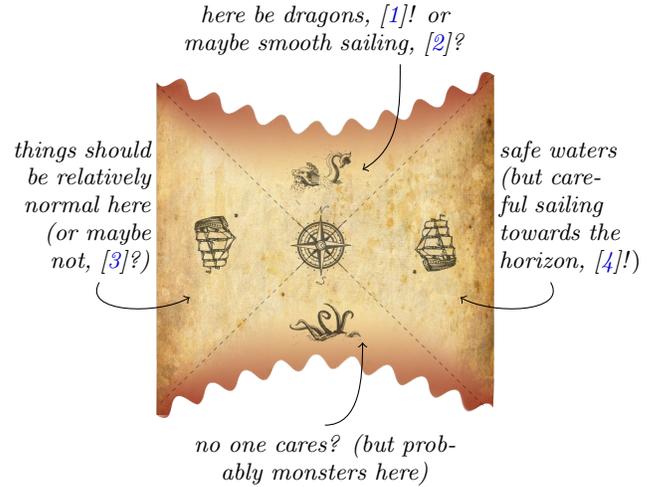
\begin{figure}[t]
\begin{tikzpicture}
\node[inner sep=0pt] at (0,0) {\includegraphics[width=0.25\textwidth]{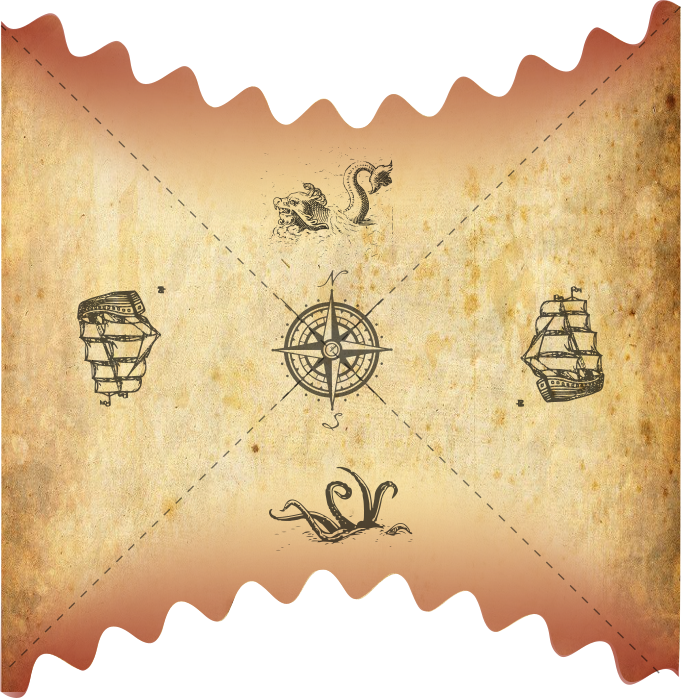}};
\draw[->] (1,2.4) to [out=-90, in=30] (0.5,1.0);
\node[above, text width=4cm, align=center] at (0,2.4) {\textit{here be dragons, \cite{Almheiri:2012rt}! or maybe smooth sailing, \cite{Papadodimas:2013wnh}?}};
\draw[->] (-3.0,-0.5) to [out=-120, in=-150] (-1.8,-0.7);
\node[left, text width=2cm, align=right] at (-2.2,0.5) {\textit{things should be relatively normal here (or maybe not, \cite{deBoer:2019kyr}?)}};
\draw[->] (3.0,-0.5) to [out=-60, in=-30] (1.8,-0.7);
\node[right, text width=2cm, align=left] at (2.2,0.5) {\textit{safe waters (but careful sailing towards the horizon, \cite{Ghosh:2017pel}!})};
\draw[->] (0,-2.4) to [out=0, in=-90] (0.5,-1.3);
\node[below, text width=4cm, align=center] at (0,-2.4) {\textit{no one cares? (but probably monsters here)}};
\end{tikzpicture}
\centering
\caption{\textit{The navigation chart of an eternal AdS black hole.}\label{fig:bh1}}
\end{figure}

The structure of quantum black holes is a perpetual source of confusion and inspiration. Any quantum gravity proposal  must explain how information leaks through the horizon of a radiating black hole. So many ideas have been put forth that the field has become difficult to navigate, see Figure \ref{fig:bh1}. The \emph{most persistent assumption}, permeating hundreds of proposals, is the idea that the Hilbert space of states of a radiating black hole \emph{splits into the tensor product} of the modes associated with the outgoing Hawking radiation and their interior partners.   In its essence, the split is a consequence of applying the equivalence principle when the Hawking radiation is derived as the Unruh radiation for an observer orbiting a black hole. Despite the fact that the validity of the equivalence principle in the presence of horizons has been repeatedly questioned, \textit{e.g.} \cite{Almheiri:2012rt}, breaking the spell of the tensor product is rarely considered.

\begin{figure}[t]
\begin{tikzpicture}[scale=2.1,xscale=-1]
\draw[black, fill=orange, opacity=0.30] (1,1.3) -- (-1,-0.7) -- (-1,-1.3) -- (1,0.7) -- cycle;
\draw[draw=black,snake it] (-1,1) to [out=-20,in=200] (1,1.3);
\draw[draw=black,snake it] (-1,-1.3) to [out=20,in=160] (1,-1);
\draw[thick, green!60!black] (-1,1) -- (-1,-1.3);
\draw[thick, green!60!black] (1,0.7) -- (1,-1);
\draw[thick, red] (1,1.3) -- (1,0.7);
\draw[dotted] (-1,1) -- (1,-1);
\draw[dotted] (-1,-0.7) -- (-0.55,-1.15);
\draw[dotted] ( 1, 0.7) -- ( 0.55, 1.15);
\draw[blue, fill=blue, opacity=0.3] (-1.2,-1.3) to [out=90, in=-90] (-1.55,-1.15) to [out=90, in=-90] (-1.2,-1.00) to (-1.2,1.0) to (-1.2,-1.3);
\draw[draw=black,snake it] (-0.4,-0.4) -- (0.3,0.3);
\draw[->] (0.3,0.3) -- (0.4,0.4);
\draw[blue, fill=blue, opacity=0.3] (1.2,1.3) to [out=-90, in=90] (1.4,1.1) to [out=-90, in=90] (1.3,0.9) to [out=-90, in=90] (1.2,0.7) to (1.2,-1.0) to (1.2,1.3);
\node[above] at (-1.4,-1.0) {$w$};
\node[below] at ( 1.4, 0.8) {$\tilde{w}$};
\node[above] at (-0.6,-0.7) {$\Phi_{\text{wh}}$};
\end{tikzpicture}
\centering
\caption{\textit{A two-sided, traversable, asymptotically AdS wormhole. While the green parts of the boundary can hold arbitrary wave packets, the boundary data $\tilde{w}$ on the red piece is uniquely determined from $w$ by the bulk evolution. }\label{fig:wh1}}
\end{figure}
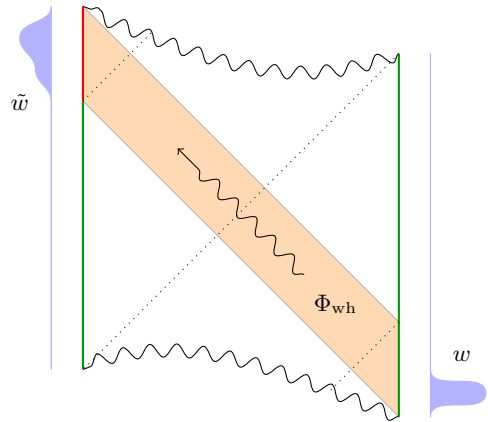

In this short essay I will entertain the preposterous idea that -- as far as radiating black holes are concerned -- such a split \emph{does not occur}. I will begin by considering \emph{wormholes} and argue the following:
\begin{claim}
Let $\mathcal{H}$ denote the Hilbert space of states dual to a two-sided, traversable, holographic wormhole, and let $\mathcal{H}_L$ and $\mathcal{H}_R$ be the Hilbert spaces of the \emph{decoupled} dual theories living on its asymptotic boundaries. In general, $\mathcal{H}_L \otimes \mathcal{H}_R \neq \mathcal{H}$.
\end{claim}
By  wormhole we mean a spacetime which has two asymptotically AdS boundaries and allows a signal to pass from one boundary to the other, as shown in Figure \ref{fig:wh1}. At this point, we do not care \textit{how} the wormhole was opened, we simply assume its existence and traversability for at least a few bits.

Consider a scalar field $\Phi$  in the bulk of the wormhole with normalizable boundary conditions. We can define its \emph{boundary limits}  as $f_{R,L} = \lim_{z_{R,L} \rightarrow 0} z_{R,L}^{-\Delta} \Phi$, where $z_{R,L}$ are the Fefferman-Graham coordinates associated with the two boundaries at $z_{R,L} = 0$ and $\Delta$ is the dimension of the dual operators. Note that $f_{R,L}$ represent the state of the theory, not the deformation. Furthermore, by $\Phi[f_L, f_R]$ let us denote any bulk field with the boundary data $f_L \times f_R$.

Holography implies that the Hilbert spaces of the quantized bulk and boundary theories match. This is the starting point of proposals such as the HKLL reconstruction \cite{Hamilton:2006az}, where  bulk operators are constructed from boundary operators. Here, we will invert this logic. We start by decomposing the bulk field into a complete set of bulk modes, $\Phi = \sum_F \Phi_F$, where $\Phi_F = \phi_F a_F + \phi_F^{\ast} a_F^{\dagger}$. Next, the coefficients $a_F^{\dagger}$ and $a_F$ are promoted to creation-annihilation operators, and the Fock space $\mathcal{H}$ is constructed. The vacuum state $| \Omega \>$ is annihilated by all $a_F$'s, and the one-particle states are $| \Phi_F \> = \Phi_F | \Omega \>$. Finally, the \emph{boundary operators} are defined by taking the boundary limits of $\Phi$.

One can see that the Hilbert space $\mathcal{H}$ splits into the tensor product $\mathcal{H}_L \otimes \mathcal{H}_R$ if and only if the boundary conditions on the left and right boundaries are independent, \textit{i.e.}, if the map $f_L \times f_R \mapsto \Phi[f_L, f_R]$ is well-defined for any boundary data. This is certainly the case for the semiclassical eternal AdS black hole. In particular, by taking the boundary limits of the quantum bulk operators $\Phi[0, f]$ and $\Phi[f, 0]$ we obtain two boundary operators $\O^{R,L}_f$. These are the usual operators smeared over each boundary, $\O^{R,L}_f = \int \D^d x \, f(x) \O^{R,L}(x)$. They act on the separate factors in the tensor product and create factorized one-particle states, $| \O^{R,L}_f  \>$.

However, the situation is different for traversable wormholes. Consider a wave packet $w$ located at the right boundary, as in Figure \ref{fig:wh1}, which travels through the wormhole and emerges on its left boundary as the wave packet $\tilde{w}$. Uniqueness of the bulk solution implies that no bulk field with the boundary data $0 \times w$ nor $\tilde{w} \times 0$ exists. We conclude that the \emph{Hilbert space does not split into the tensor product}.

One might worry that this set-up suffers from issues such as a lack of coupling to gravity, the wormhole collapsing under the incoming wave, \textit{etc}. Notice, however, that the essence of the argument relies only on the fact that some field configurations, created at one boundary, hit the opposite boundary. The precise mechanism is irrelevant. Furthermore, the argument does not depend on the choice of modes in the decomposition of the bulk field nor on the choice of the vacuum.

So what are the \emph{states that violate the split} into the tensor product? Just to keep the discussion simple, consider a single \emph{wormhole mode}, $\Phi_{\text{wh}} = \Phi[\tilde{w}, w]$, which traverses the wormhole. The states we are looking for, $| \O^R_w \>$ and $| \O^L_{\tilde{w}} \>$, are the \emph{would-be} one-particle states associated with the boundary data carried by the wormhole mode. They can be defined using the boundary limits of any bulk fields $\Phi[f_L, w]$ and $\Phi[\tilde{w}, f_R]$ for $f_L$ and $f_R$ \emph{consistent} with $w$ and $\tilde{w}$ on opposite boundaries. However, since we cannot set $f_L = f_R = 0$, we conclude:
\begin{claim}
In the \emph{semiclassical approximation}, where the split into the tensor product is assumed, we would expect that $| \O^R_w \>$ belongs to $\mathcal{H}_R$, while $| \O^L_{\tilde{w}} \>$ is an element of $\mathcal{H}_L$. But in reality, the two states are linearly dependent, as they are the boundary limits of the single state $\Phi_{\text{wh}} | \Omega \> = \phi_{\text{wh}}^{\ast} a_{\text{wh}}^{\dagger} | \Omega \>$.
\end{claim}

We can now address the question of how the non-trivial structure of the Hilbert space \emph{manifests itself in the semiclassical approximation}. Since we would expect that the states $|\O^R_w \>$ and $| \O^L_{\tilde{w}} \>$ are different, their non-vanishing scalar product $\< \O^L_{\tilde{w}} | \O^R_w \>$ could be interpreted as the presence of an interaction.

\begin{claim}
In the semiclassical approximation, the departure of the Hilbert space $\mathcal{H}$ from the tensor product $\mathcal{H}_L \otimes \mathcal{H}_R$ manifests itself as the \emph{emergent, non-local interaction} $H_{\text{int}} = \O^L_{\tilde{w}} \O^R_w$.
\end{claim}
In the seminal work of Gao, Jafferis, and Wall, \cite{Gao:2016bin}, a wormhole was opened by introducing a coupling between the QFTs living on the boundaries of the BTZ black hole. As we can see, this \emph{non-local interaction} between the two QFTs, which sustains the wormhole, \emph{is emergent}. It is simply the \emph{avatar of the non-trivial structure of the Hilbert space of states} describing the holographic wormhole.

Using our new-found knowledge of wormholes, let us now speculate about the structure of the Hilbert spaces of radiating black holes. From the early days, we know that in order to unitarize  black hole evolution, the interior modes must be somehow encoded in the Hawking radiation. And after decades of research, what is generally accepted today is that, in Susskind's words, \cite{Susskind:2014moa}, \emph{`Entanglement is not enough}.' There exists some mysterious \emph{long-range, non-local interaction} between the degrees of freedom on both sides of the horizon. Based on the wormhole story, is it possible that what we are actually looking for is a \emph{change in the structure of the Hilbert space} $\mathcal{H}_L \otimes \mathcal{H}_R$ of the semiclassical black hole rather than a genuine interaction?

\begin{figure}[t]
\includegraphics[width=0.40\textwidth]{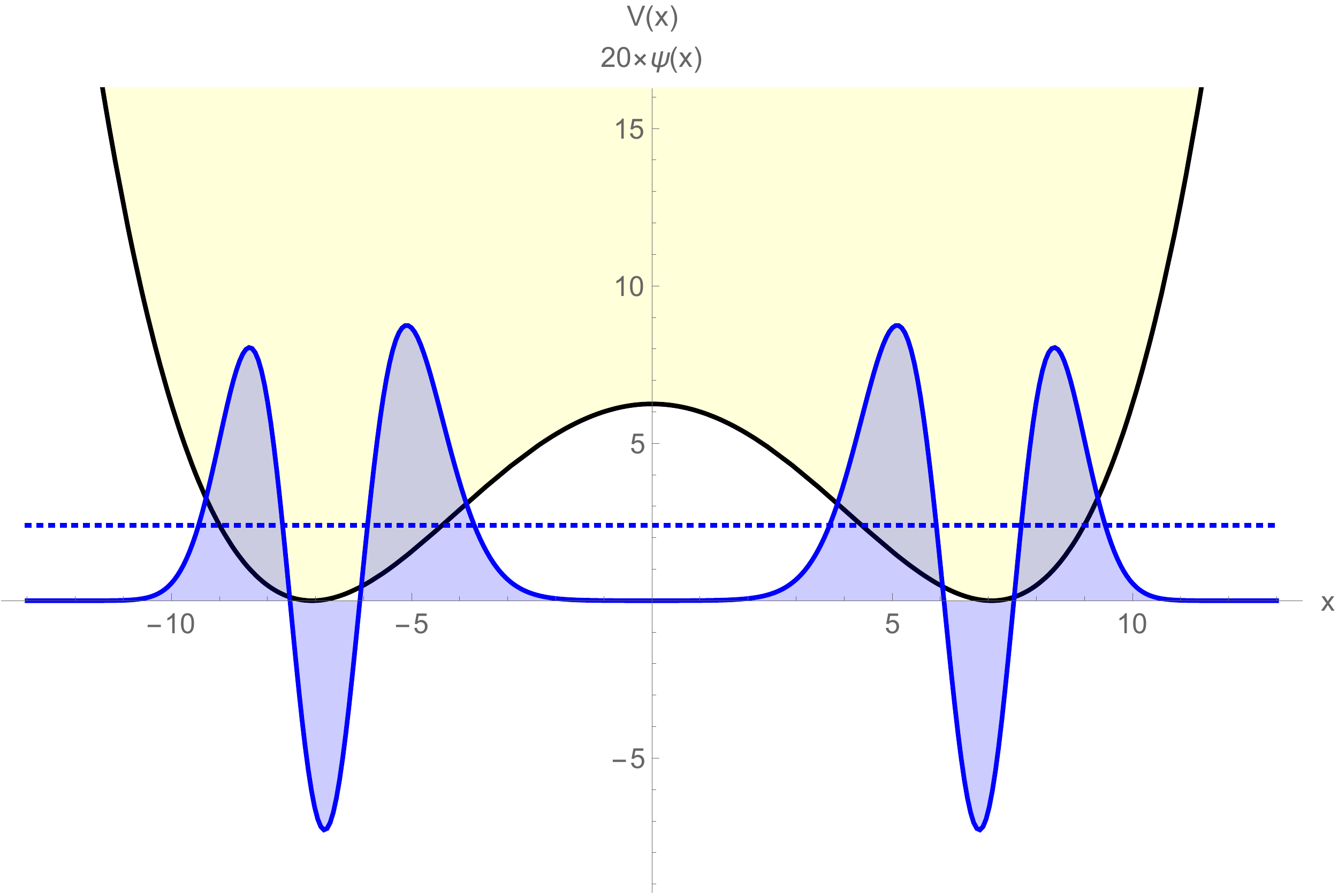}
\caption{\textit{In a 0+1-dimensional model, the two related modes, $\O^L_{\omega}$ and $\O^R_{\omega}$, describe small excitations around the two minima of the potential. The plot shows the wave function of a symmetric energy eigenstate with an energy value well below the potential barrier, indicated by the dotted line. Due to the high potential barrier, the excitations around the left and right minima appear almost independent and the Hilbert space effectively splits into the tensor product. See \cite{Bzowski:2018aiq} for a detailed discussion.}\label{fig:model1}}
\end{figure}

As an oversimplified example, consider a pair of entangled modes on both sides of the horizon. The key observation is that in order to introduce a coupling  we do not need to take the tensor product of their Hilbert spaces. Instead, the modes are coupled in the way two approximate harmonic oscillators are coupled in the \emph{double-well system}, see Figure \ref{fig:model1}. The excitations around the two minima corresponds to semiclassical physics close to the asymptotic boundaries, the tunneling effects are interpreted as Hawking radiation, as in \cite{Parikh:1999mf}, the dynamics of the inverted harmonic oscillator close to the tip of the potential barrier governs the near-horizon scattering processes, as found in \cite{Betzios:2016yaq,Hashimoto:2020xfr}, and the \emph{binding energy} of the true ground state is interpreted as the negative energy required for sustaining the wormhole.

The double-well construction introduces a coupling, which leads to both perturbative and non-perturbative effects. However, in the case of the black hole, there exist models where the physics in the left and right wedges is modified purely by non-perturbative effects and thus is indistinguishable from the original black hole in the semiclassical approximation. These include models based on various identifications (\textit{e.g.}, the antipodal identification, \cite{Hooft:2016vug}) or additional boundary conditions (\textit{e.g.}, end-of-the-world branes, \cite{Rozali:2019day}). Such additional conditions relate  bulk field configurations between the left and right wedges of the black hole. While the evolution of the entire system remains causal, a boundary observer perceives non-perturbatively small non-localities, which grow when probing regions closer to the horizon.

To summarize, in this short essay I argued that the Hilbert space of states dual to a wormhole or a radiating black hole does not split into the tensor product, $\mathcal{H}_L \otimes \mathcal{H}_R$, of its boundary theories. Although this possibility was hinted at in \textit{e.g.}, \cite{Harlow:2018tqv}, in this essay I have outlined a general proof. The consequences of this observation raise a number of questions regarding the validity of results obtained under the spell of the tensor product. This includes any geometrical methods in the bulk which rely on the evocation of the equivalence principle beyond the horizon. Furthermore, any reasoning based on entanglement, where tracing is carried out over a factor in the tensor product, must be reexamined. In particular, to what extent can methods relying on holographic entanglement entropy, \textit{e.g.}, \cite{Almheiri:2020cfm},  capture the non-triviality of the Hilbert space of states?

\hspace{1cm}

\noindent\textit{Acknowledgments.} I would like to thank Alessandra Gnecchi, Marjorie Schillo and Panos Betzios for their help in the preparation of the essay. I am supported by the Advanced ERC grant SM-grav, No. 669288.

\end{document}